# Laser-assisted interaction between nonrelativistic electrons and positrons

S. S. Starodub[1], S. P. Roshchupkin[2] and V.V. Dubov[2]

[1]Institute of Applied Physics, National Academy of Sciences of Ukraine,
58 Petropavlovskaya Str., Sumy 40000, Ukraine
[2]Department of Theoretical Physics, Peter the Great St. Petersburg Polytechnic University,
195251, St-Petersburg, Russian Federation, Russia
E-mail: starodubss@gmail.com, serg9rsp@gmail.com, maildu@mail.ru

The effective interaction between two classical nonrelativistic electrons (positrons) in the presence of intense electromagnetic radiation (one and two waves) is theoretically studied. Small relativistic corrections are taking into account in the laboratory reference frame. The field of an intense wave forms the movement of particles in such a way that their trajectories are practically parallel for quite a long time, and in the perpendicular direction the particles shift slightly, approaching first and then moving away from each other. This result can be considered as an effective attraction of same charged particles. Shown that the effective attraction of two electrons and two positrons could be substantially asymmetric.

## 1. Introduction

Interest in the interaction of same charged particles in an external electromagnetic field begins in the second half of the 20th century (see the review [1]). Oleinik firstly showed the possibility of electrons attraction due to their interaction with a plane electromagnetic wave in 1967 [2]. However, A. Kazantsev and V. Sokolov provided a theoretical proof for such an effect only in 1984 [3] for classical relativistic electrons in the field of a plane monochromatic electromagnetic wave. They found that for very intense radiation the effective potential can, in fact, be attractive and may lead to the aforementioned bound states. S. Zavtrak (1989) has investigated the force of radiation pressure in relation to the Coulomb force. [4]. He showed the principle possibility of the formation of bound states between same charged particles. It is very important to point out that the attraction of classical electrons in the field of a plane monochromatic electromagnetic wave is possible only for particles with relativistic energies. In the last decade, experiments on the interaction of a matter with an electromagnetic field are carried out using ultrashort and very powerful laser pulses. [5]. Experiments with lasers, whose peak intensity of the order of $10^{21}\ \text{W}/\text{cm}^2$ is becoming common practice [6]. Moreover, this is not the limit. Projects where the peak intensity of lasers will be about $10^{23}\ \text{W}/\text{cm}^2$ are realizing today [ELI, XCELS]. In such powerful fields, new nonlinear electrodynamic effects will appear both classical and quantum. Earlier in a series of papers, Starodub, Roshchupkin and co-authors [7–10] studied the effective forces experienced by two electrons (e-e interaction) and the same charged ions interaction with pulses electromagnetic radiation in the center-of-mass system. Thus, in the review [7] the following processes were discussed: nonrelativistic e-e interaction (and light ions) in the pulsed field of a single laser wave; e-e interaction in the pulsed field of two counter-propagating laser waves moving perpendicularly to the initial direction of electrons motion; the interaction of nonrelativistic light ions moving almost parallel to each other in the propagation direction of the pulsed field of two counter-propagating laser waves moving in parallel direction to ions; interaction of two nonrelativistic heavy nuclei (uranium 235), moving towards each other perpendicularly to the propagation direction of two counter-propagating laser waves. Influence of pulsed field of two co-propagating laser waves on the effective force of e-e interaction and two identically charged heavy nuclei was studied in [8]. It was shown that the phase shift allows to increase duration of electron's confinement at a certain averaged effective distance by 1.5 time in comparison with the case of one and two counter-propagating pulsed laser waves. Interaction of two classical non-relativistic electrons in the strong-pulsed laser field of two light mutually perpendicular waves, when the maxima laser pulses coincide, was studied in Starodub et al. [9]. It is shown that the effective force of electron interaction becoming the attraction force or anomalous repulsion force after approach of electrons to the minimum distance. In [10], allowing for the phase shifts of pulse peaks made it possible to significantly change the effective interaction of electrons. It is important to emphasize that in the works of the authors the interaction of electrons (positrons and ions) was studied in the center-of-mass system with negligible small corrections. As a result, the motion of the center of mass and the relative motion of the particles were considered independently of each other.

In the present paper, in contrast to the mentioned above, interaction of identical non-relativistic electrons (e-e interaction) and positrons (p-p interaction) in the inertial reference system is considered. It was taking into account all the necessary relativistic corrections (in the center of mass system the center of mass movement and relative particles motion are not separated).

The obtained results can be used for experiments in the framework of modern research projects, where the sources of pulsed laser radiation are used SLAC, ELI, XCELS [6, 11-12].

**2. Theoretical part**

Our concept (see Fig.1) is based on the opposite movement of two non- relativistic electrons in the field of one or two pulse laser waves. In the beginning, particles are on the $x$ axis. Two linearly polarized electromagnetic waves are perpendicular. The first wave propagates along the $z$ axis, the second wave propagates along the $x$ axis (on the same axis as the particles).

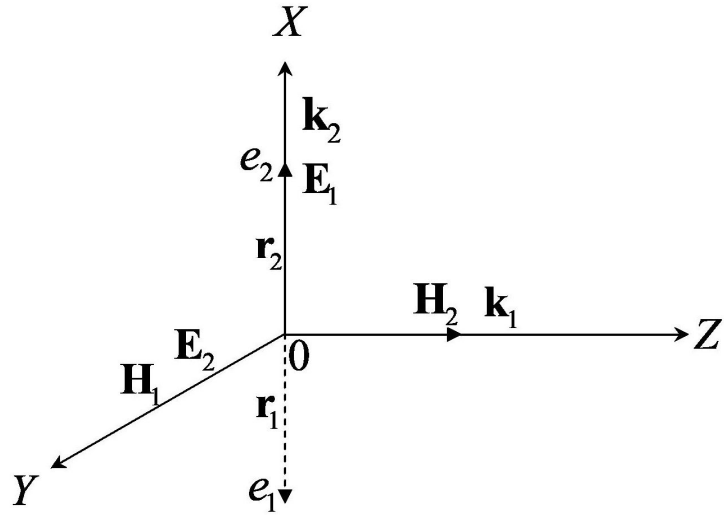

**Figure 1.** Interaction kinematics of two classical particles with point charge $e_{1,2}$ in the field of two light mutually perpendicular waves.

The strengths of the electric and magnetic fields are given in the following form:

$$\mathbf{E}(t, z_j, x_j) = \mathbf{E}_1(t, z_j) + \mathbf{E}_2(t, x_j). \tag{1}$$

$$\mathbf{E}_1(t, z_j) = E_{01} \cdot \exp\left[-\left(\frac{\varphi_{1j}}{\omega_1 t_1}\right)^2\right] \cos\varphi_{1j} \cdot \mathbf{e}_x, \quad \varphi_{1j} = (\omega_1 t - k_1 z_j), \tag{2}$$

$$\mathbf{E}_2(t, x_j) = E_{02} \cdot \exp\left[-\left(\frac{\varphi_{2j}}{\omega_2 t_2}\right)^2\right] \cos\varphi_{2j} \cdot \mathbf{e}_y, \quad \varphi_{2j} = (\omega_2 t - k_2 x_j). \tag{3}$$

$$\mathbf{H}(t, z_j, x_j) = \mathbf{H}_1(t, z_j) + \mathbf{H}_2(t, x_j). \tag{4}$$

$$\mathbf{H}_1(t, z_j) = H_{01} \cdot \exp\left[-\left(\frac{\varphi_{1j}}{\omega_1 t_1}\right)^2\right] \cos\varphi_{1j} \cdot \mathbf{e}_y, \tag{5}$$

$$\mathbf{H}_2(t, x_j) = H_{02} \cdot \exp\left[-\left(\frac{\varphi_{2j}}{\omega_2 t_2}\right)^2\right] \cos\varphi_{2j} \cdot \mathbf{e}_z. \tag{6}$$

Hereinafter, index $j = 1, 2$ denotes the first and second particles, and index $i = 1, 2$ denotes the number of the wave; $\varphi_{ij}$ - the phases of the corresponding waves and particles; $E_{0i}$ and $H_{0i}$ are the strength of the electric and magnetic field in the pulse peak, respectively; $t_i$ and $\omega_i$ are the pulse duration and frequency of the waves; $\mathbf{e}_x$, $\mathbf{e}_y$, $\mathbf{e}_z$ are unit vectors directed along the $x$, $y$, $z$ axes.

Further we assume the corresponding amplitudes of intensities of electric and magnetic fields in a plane wave are equal among themselves $(E_{01} = H_{01}, E_{02} = H_{02})$, and the frequencies are equal too: $\omega_1 = \omega_2 = \omega$, $|\mathbf{k}_1| = |\mathbf{k}_2| = k = \omega/c = \lambdabar^{-1}$.

We treat the electron interaction classically, thus Newton equations for motion of two identically charged particles with the mass $m$ and charge $e$ ($e = e_1 = e_2$) in the pulsed field of two mutually perpendicular laser waves and taking into account relativistic corrections of the order $v/c \ll 1$ in the inertial reference system are determined by the following expressions:

$$m\ddot{\mathbf{r}}_{1(\mp)} = \mp|e|\left[\mathbf{E}(t, z_{1(\mp)}, x_{1(\mp)}) + \frac{\dot{\mathbf{r}}_{1(\mp)}}{c} \times \mathbf{H}(t, z_{1(\mp)}, x_{1(\mp)})\right] - \frac{e^2(\mathbf{r}_{2(\mp)} - \mathbf{r}_{1(\mp)})}{|\mathbf{r}_{2(\mp)} - \mathbf{r}_{1(\mp)}|^3}, \quad (7)$$

$$m\ddot{\mathbf{r}}_{2(\mp)} = \mp|e|\left[\mathbf{E}(t, z_{2(\mp)}, x_{2(\mp)}) + \frac{\dot{\mathbf{r}}_{1(\mp)}}{c} \times \mathbf{H}(t, z_{2(\mp)}, x_{2(\mp)})\right] + \frac{e^2(\mathbf{r}_{2(\mp)} - \mathbf{r}_{1(\mp)})}{|\mathbf{r}_{2(\mp)} - \mathbf{r}_{1(\mp)}|^3}, \quad (8)$$

here $\mathbf{r}_{1(\mp)}$ and $\mathbf{r}_{2(\mp)}$ are electron (negative sign "–") or positrons (positive sign "+") radius-vectors. Further, to simplify the formula, let's omit the sign $\mp$ in the designations at the radius-vectors and their components, but leave it at the charge.

After performing simple transformations and transition to dimensionless values, the equations of electrons (positrons) motion will take the form:

$$\ddot{\xi}_1 = \mp(D_{1x}\mathbf{e}_x + D_{1y}\mathbf{e}_y + D_{1z}\mathbf{e}_z) - \beta\frac{(\xi_2 - \xi_1)}{|\xi_2 - \xi_1|^3}, \quad (9)$$

$$\ddot{\xi}_2 = \mp(D_{2x}\mathbf{e}_x + D_{2y}\mathbf{e}_y + D_{2z}\mathbf{e}_z) + \beta\frac{(\xi_2 - \xi_1)}{|\xi_2 - \xi_1|^3}, \quad (10)$$

here:

$$D_{jx} = \eta_1 f_{1j}(1 - \dot{\xi}_{jz})\cos(\tau - \xi_{jz}) + \eta_2 f_{2j}\dot{\xi}_{jy}\cos(\tau - \xi_{jx}), \quad (11)$$

$$D_{jy} = \eta_2 f_{2j}(1 - \dot{\xi}_{jx})\cos(\tau - \xi_{jx}), \quad D_{jz} = \eta_1 f_{1j}\dot{\xi}_{jx}\cos(\tau - \xi_{jz}), \quad j = 1, 2. \quad (12)$$

Where $\tau = \omega t$, $\xi_j = k\mathbf{r}_j = (\mathbf{r}_j/\lambdabar)$, $\dot{\xi}_j = (d\mathbf{r}_j/d\tau)$, $\ddot{\xi}_j = (d^2\mathbf{r}_j/d\tau^2)$,

$$f_{1j} = \exp\left(-\frac{(\tau - \xi_{jz})^2}{\tau_1^2}\right), \quad f_{2j} = \exp\left(-\frac{(\tau - \xi_{jx})^2}{\tau_2^2}\right). \quad (13)$$

$$\beta = \frac{(e^2/\lambdabar)}{mc^2}, \quad \eta_i = \frac{eE_{0i}\lambdabar}{mc^2}, \quad i = 1, 2. \quad (14)$$

In (13)-(14) $\tau_i$ - are first and second wave pulse durations, $\xi_j$ are electrons (positrons) radius-vectors in units of wave length, the classical parameters $\eta_i$ are numerically equal to the ratio of the oscillation velocity of a charge particle in the peak of a pulse of the first or second wave to the velocity of light $c$ (hereinafter, should consider parameters $\eta_j$ as oscillation velocities); the parameter $\beta$ is numerically equal to the ratio of the energy of Coulomb interaction of particles with charge $e$ at the wavelength to the particles rest energy.

Equations (9)-(14) exactly take into account the interaction of electrons (positrons) with each other, as well as with the field of electromagnetic waves and do not have an analytical solution in the non-relativistic energy limit. Therefore, this task we solved numerically. The initial conditions of the described model are as follows: the distance between the electrons is $\xi = \xi_2 - \xi_1 = 2$ ($\xi_1 = -1$, $\xi_2 = 1$) in units of or 200nm; wavelength is $\lambdabar = 100$nm ($\omega = 3 \times 10^{15} \text{s}^{-1}$). The initial velocity of electrons is $\dot{\xi}_1 = 7 \times 10^{-4}$, $\dot{\xi}_2 = -7 \times 10^{-4}$, which corresponds to the initial kinetic energy of 0.1eV. The interaction time (time for the particles to approach and scatter to the initial distance without an external field see Fig.

2a) is $\tau = 3000$ ($t = 1\text{ps}$). This parameter increased during the simulation process when it was necessary. Pulse duration and oscillation velocities (wave intensities) varied during the simulation process.

## 3. Results of numerical modeling

Without external field the same charged particles interact according to the Coulomb Law (see Fig. 2a, b).

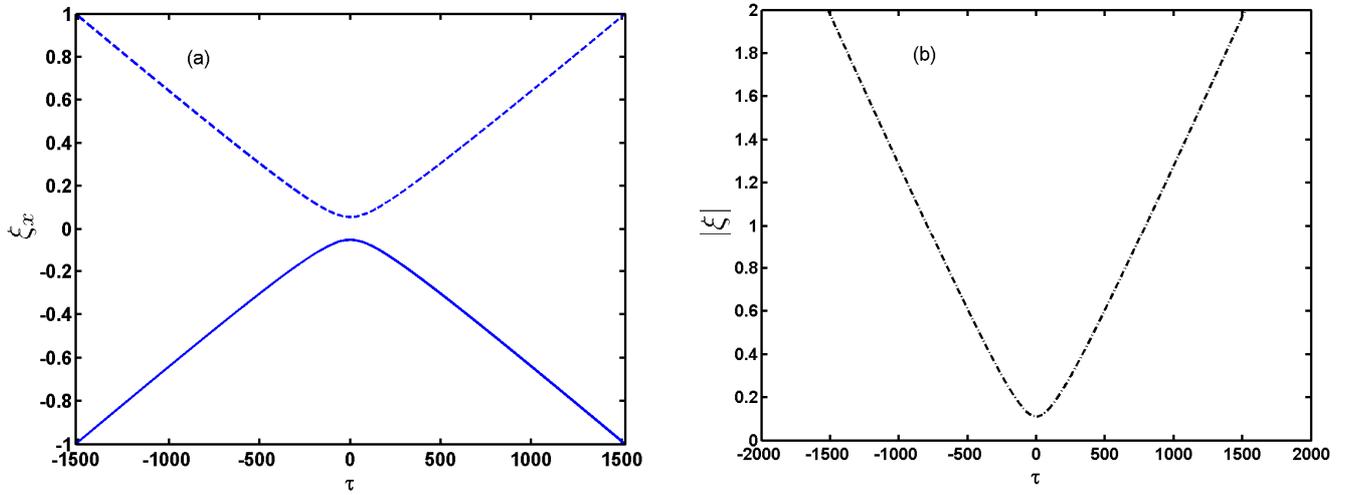

**Figure 2.** a) The trajectories of electron's (positron's) Coulomb interaction (first particle - blue dotted line and second particle - blue solid line) along the axis $x$ (in units $\lambdabar$) against the interaction time $\tau$ (in units $\omega t$); b) Particles relative distance module $|\xi|$ against the interaction time $\tau$. External field is absent $\eta_1 = \eta_2 = 0$.

Fig 2a shows the trajectory of first electron (positron) (blue dotted line) and the trajectory of second particle (blue solid line) without external field. The ordinate axis shows the value of the distance $\xi_x$, as the other components are equal to zero due to the selected initial values. In this case the interaction time is selected in such a way that at the point $\tau = 0$ the electrons uniformly approach each other at the minimum possible distance (it corresponds to $\xi_{x\,\min} \approx 10^{-1}$ ($\approx 10\text{nm}$)). The external pulsed electromagnetic field should have a maximum intensity at this point exactly. The particles relative distance module against the interaction time is plotted in Fig2b. The plot shows uniform change in the distance between particles during the entire interaction time. Note that the electrons (positrons) are approach and scatter during the time $\tau = 3000$.

The interaction of two electrons (blue lines) and two positrons (red lines) in the field of a plane pulsed wave propagating along the axis $z$ is shown in Figure 3.

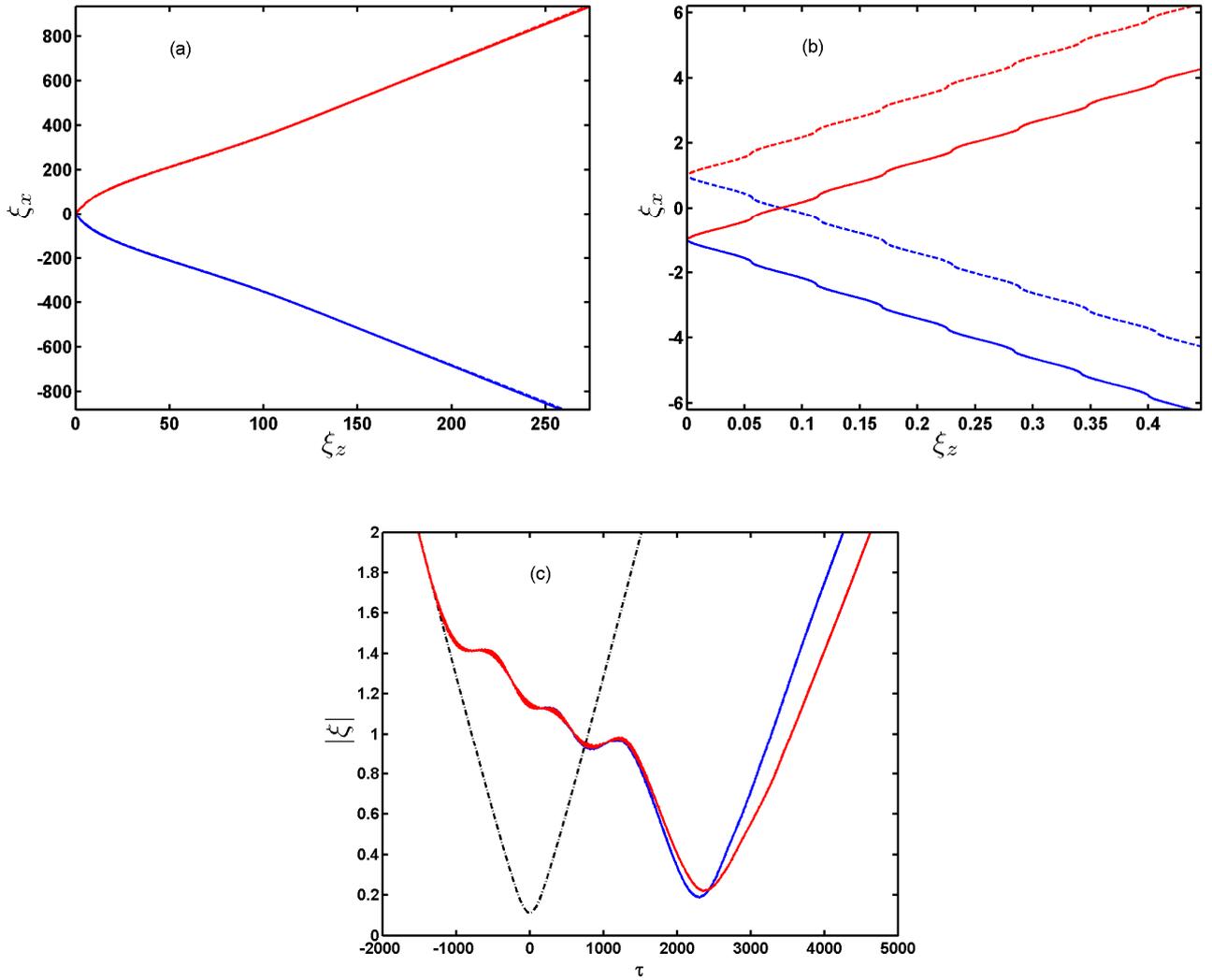

**Figure 3.** Trajectories of electrons (blue line) and positrons (red line) in the $xz$ plane. a) The trajectories of two particles in full scale merge into one line, b) Small change region $\xi_x$, $\xi_z$ at the beginning of particle interaction (enlarged scale of the Fig. 2a): dotted blue line – first electron, dotted red line – first positron. c) Particles relative distance module $|\xi|$ against the interaction time $\tau$, dashed-dotted black line mark Coulomb interaction. External field parameters are: $\eta_1 = 3 \times 10^{-1}$ ($I_1 \approx 3 \times 10^{17}$ W/cm$^2$), $\tau_1 = 1500$.

The field was chosen strong in order to the oscillation velocity ($\eta_1$) was much greater than the initial velocity of charged particles. It corresponds to the first wave field intensity of order-of-magnitude $I_1 \Box 10^{17}$ W/cm$^2$. The pulse duration is $\tau_1 = 1500$ ($t_1 = 0.5$ps). A strong field forms the motion of electrons (positrons) in such a way that their trajectories on a large scale are practically parallel to each other (see Fig. 3a) and Fig.3b). The plots differ only in the scale of the particle coordinates. Figure 2a) shows that the charged particles cover a distance greater than their initial one up to 450 times due to the action of a strong field (see change of component $\xi_x$). Formation of such a long, practically parallel beam of two particles becomes possible due to the choice of the initial phase $\varphi_{ij}$ in eqs. (2), (3). It is necessary to select a phase at which the initial influence of the field will be maximal. At initial phase shift by $\pi$-radian direction of the particles trajectory change from negative area to positive and vice versa. The relative distance between electrons (positrons) changes irregularly when interacting with electromagnetic radiation of one wave. At first, the distance decreases slowly, almost by an order of magnitude (until $|\xi|_{min} \approx 2 \times 10^{-1}$, see Fig. 3c), and then the particles move away from each other quite quickly to initial distance ($|\xi| \approx 2$). Furthermore, time during which particles return to an initial relative distance increases up to 2.8 times, in comparison with corresponding time in a Coulomb field. These results indicate that in the wave field there is an effective attraction of electrons (positrons). The difference in the effective attraction between electrons and positrons is small and occurs itself only at the end of the interaction.

The next case is the influence of a second pulsed electromagnetic wave directed along the $x$-axis (perpendicular to the first wave and along the initial position of the particles) on the motion of electrons (positrons). Numerical analysis shows that the intensity of the second wave and pulse duration should be much less than the corresponding parameters of the first one. In this case, an increase in the effective attraction of the same charged particles may occur in comparison with a single wave field.

The electrons (blue line) and positrons (red line) relative distance module against interaction time in the field of two pulse waves are plotted in Fig.4. In Fig.4a) the intensity of the second wave is 6 times less than in Fig.4b) and the intensity of the first one is the same in both plots.

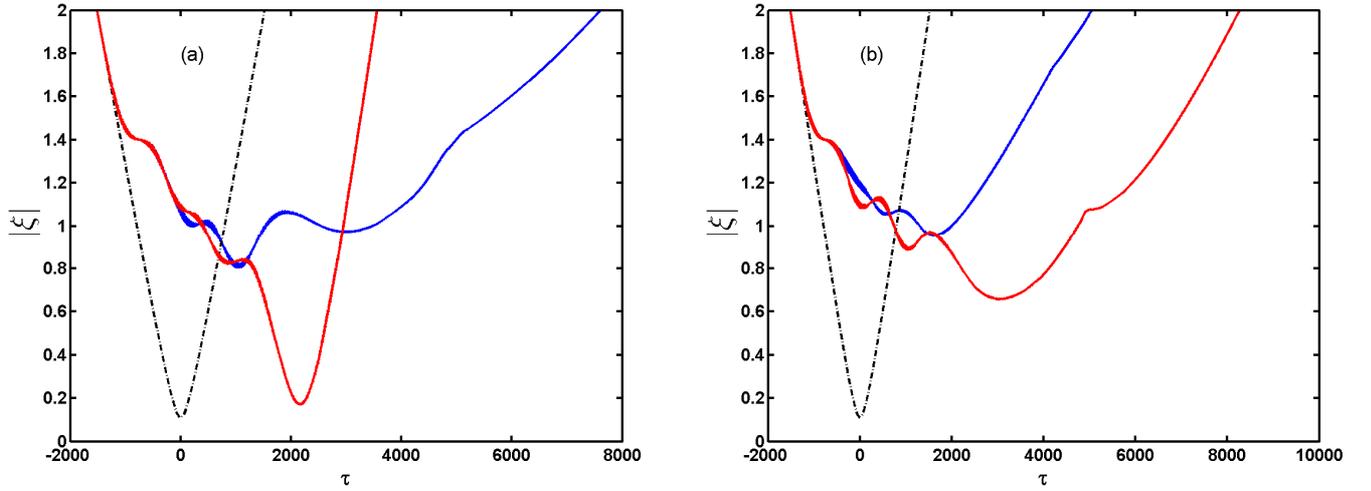

**Figure 4.** The electrons (blue line) and positrons (red line) relative distance module $|\xi|$ against the interaction time $\tau$ in a field of two pulsed wave. Here: $\eta_1 = 3\times 10^{-1}$ ($I_1 \approx 3\times 10^{17}$ W/cm$^2$), $\tau_1 = 1500$, $\tau_2 = 300$ in both plots, a) $\eta_2 = 10^{-3}$ ($I_2 = 3.4\times 10^{12}$ W/cm$^2$); b) $\eta_2 = 6\times 10^{-3}$ ($I_2 = 1.2\times 10^{14}$ W/cm$^2$); dashed-dotted black line mark Coulomb interaction.

Fig. 4a) shows that electrons return to the initial distance ($|\xi|=2$) in a time twice as long as the corresponding time for positrons ($\Delta\tau = 4053$). I.e. for these field values, the effective attraction of electrons is significantly greater than the one of positrons. If the intensity of the second wave increases by 6 times the view of effective interaction of electrons and positrons changes to the opposite (see Fig.4b): the effective attraction of positrons becomes greater than the electrons (up to 1,6 times, $\Delta\tau = 3219$). Note that Figs. 4a), b) show the greatest influence of both waves on the effective interaction of electrons (positrons). Furthermore, it is possible to select such value of intensity of the second wave that the effective attraction between two electrons and two positrons would be practically identical.

Analysis of comparison Fig. 3c) with Figs. 4a), 4b) shows that in the field of two waves, both electrons and positrons scatter by initial distances ($|\xi| \approx 2$) up to 1.6 times (in units $\tau$) longer than in the field of one wave and up to 3 times longer than without an external field (see Fig.2b). Thus, the second wave, directed perpendicular to the first wave and along the initial motion of the particles, may affect to the particles velocity. In this case, the particle trajectories also lengthen (see Fig.5).

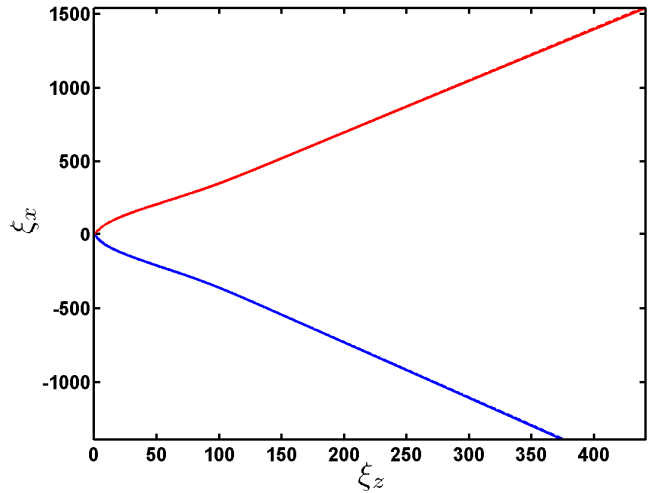

**Figure 5.** Trajectories of electrons (blue line) and positrons (red line) in the $xz$ plane. $\eta_1 = 3\times10^{-1}$ ($I_1 \approx 3\times10^{17}$ W/cm$^2$), $\tau_1 = 1500$, $\tau_2 = 300$ Solid blue line: $\eta_2 = 10^{-3}$ ($I_2 = 3.4\times10^{12}$ W/cm$^2$); solid red line: $\eta_2 = 6\times10^{-3}$ ($I_2 = 1.2\times10^{14}$ W/cm$^2$).

Fig. 5 shows the trajectories of electrons (blue line) and positrons (red line) in the pulsed field of two waves. The trajectory of electrons is plotted according to the intensities in Fug.4a). The trajectory of positrons is plotted according to the intensities in Fug.4b). Absolute final values of the particle coordinates differ insignificantly, but exceed the corresponding values in the field of one wave (compare Fig. 3a) and Fig.5). On average, both positrons and electrons scatter up to 1.6 times far along the $x$ and $z$ axes in contrast with single wave field. However, the relative distance remains the same. Thus, the field of two pulsed waves further enhances the retention of same charged particles in comparison with the field of one wave.

## 4. Conclusions

In conclusion, our results show that the strong wave field forms the trajectories of electrons (positrons) in such a way that they move practically in parallel to each other for a long time, first approaching and then moving away from each other by insignificant distances compared to the longitudinal motion. As a result, the time of particles return to the initial distance ($2\lambdabar$) increases significantly (up to 3 times). We predict that effective attraction of same charged particles in the presence of external pulsed electromagnetic radiation is occur. It is important to note that the effective interaction of two electrons and two positrons can be essentially asymmetrical. Changing the intensity of the second wave allows increasing (decreasing) the effective attraction of electrons in comparison with the corresponding interaction of positrons.